\documentclass{desyproc}

\begin{document}
\title{Searching for Scalar Dark Matter in Atoms and Astrophysical Phenomena:~Variation of \\ Fundamental Constants}

\author{{\slshape Yevgeny V.~Stadnik$^1$, Benjamin M.~Roberts$^{1}$, Victor V.~Flambaum$^{1,2}$, Vladimir A.~Dzuba$^1$}\\[1ex]
$^1$ School of Physics, University of New South Wales, Sydney 2052, Australia\\
$^2$ Mainz Institute for Theoretical Physics, Johannes Gutenberg University Mainz, D 55122 Mainz, Germany}

\contribID{Roberts\_Benjamin\_Scalars}

\confID{11832}  
\desyproc{DESY-PROC-2015-02}
\acronym{Patras 2015} 
\doi  

\maketitle

\begin{abstract}
We propose to search for scalar dark matter via its effects on the electromagnetic fine-structure constant and particle masses. Scalar dark matter that forms an oscillating classical field produces `slow' linear-in-time drifts and oscillating variations of the fundamental constants, while scalar dark matter that forms topological defects produces transient-in-time variations of the constants of Nature. These variations can be sought for with atomic clock, laser interferometer and pulsar timing measurements. Atomic spectroscopy and Big Bang nucleosynthesis measurements already give improved bounds on the quadratic interaction parameters of scalar dark matter with the photon, electron, and light quarks by up to 15 orders of magnitude, while Big Bang nucleosynthesis measurements provide the first such constraints on the interaction parameters of scalar dark matter with the massive vector bosons.

\end{abstract}

\section{Introduction}
Astrophysical observations indicate that the matter content of the Universe is overwhelmingly dominated by dark matter (DM), the energy density of which exceeds that of ordinary matter by a factor of five. In order to explain the observed abundance of DM, it is reasonable to expect that DM interacts non-gravitationally with ordinary matter. Searches for weakly interacting massive particle (WIMP) DM, which look for the scattering of WIMPs off nuclei, have not yet produced a strong positive result. Further progress with these traditional searches is hindered by the observation that the sought effects are fourth-power in the underlying interaction strength between DM and Standard Model (SM) matter, which is known to be extremely small. 

We propose to search for other well-motivated DM candidates that include ultralight (sub-eV mass) scalar particles, which are closely related to axions (the only difference being the intrinsic parity of the particle, which determines the forms of the interactions with SM particles), by exploiting effects that are first-power in the interaction strength between these scalar particles and SM matter. Ultralight scalar (and axion-like pseudoscalar) DM in the mass range $m_\phi \sim 10^{-24} - 10^{-20}$ eV has been proposed \cite{Gruzinov2000fuzzyDM,Marsh2013DMHalo,Schive2014B} to solve several long-standing astrophysical puzzles, such as the cusp-core, missing satellite, and too-big-to-fail problems \cite{Weinberg2015ReviewDMIssues}, due to its effects on structure formation. Scalar DM can produce variations of the fundamental constants of Nature, which leave distinctive signatures in atomic clock, laser interferometer and pulsar timing observables. The phenomenally high level of precision already attainable in measurements performed with these systems provides strong motivation to utilise them in searches for scalar DM.

\section{Scalar dark matter}
Scalar (as well as axion-like pseudoscalar) DM may interact quadratically with SM matter as follows:
\begin{align}
\label{Scalar_couplings_quad}
\mathcal{L}_{\textrm{int}} = \frac{\phi^2}{(\Lambda'_\gamma)^2} \frac{F_{\mu \nu} F^{\mu \nu}}{4} - \sum_{f} \frac{\phi^2}{(\Lambda'_f)^2} m_f \bar{f} f + \sum_{V} \frac{\phi^2}{(\Lambda'_V)^2} \frac{M_V^2}{2} V_\nu V^\nu ,
\end{align}
where the first term represents the coupling of the scalar field to the electromagnetic field tensor $F$, the second term represents the coupling of the scalar field to the fermion bilinears $\bar{f} f$, while the third term represents the coupling of the scalar field to the massive vector boson wavefunctions. The $\Lambda'_X$ that appear in Eq.~(\ref{Scalar_couplings_quad}) are moderately large energy scales that are constrained from stellar energy-loss arguments: $\Lambda'_\gamma \gtrsim 3 \times 10^{3}~\textrm{GeV}$, $\Lambda'_p \gtrsim 15 \times 10^{3}~\textrm{GeV}$, $\Lambda'_e \gtrsim 3 \times 10^{3}~\textrm{GeV}$, which are stronger than bounds from fifth-force searches: $\Lambda'_p \gtrsim 2 \times 10^{3}~\textrm{GeV}$ 
(note that a fifth force due to the quadratic couplings in Eq.~(\ref{Scalar_couplings_quad}) arises in the leading order through the exchange of a pair of $\phi$-quanta between two fermions, which generates a less efficient $V(r) \simeq - m_f^2 / 64 \pi^3 (\Lambda'_f)^4 \cdot 1 / r^3$ attractive potential, instead of the usual Yukawa potential in the case of linear-in-$\phi$ couplings) \cite{Olive2008}. 

The couplings in Eq.~(\ref{Scalar_couplings_quad}) alter the electromagnetic fine-structure constant $\alpha$ and particle masses as follows:
\begin{align}
\label{Quad_constants}
\alpha \to \frac{\alpha}{1-\phi^2/(\Lambda'_\gamma)^2} \simeq \alpha \left[1 +\frac{\phi^2}{(\Lambda'_\gamma)^2} \right] , ~\frac{\delta m_f}{m_f} = \frac{\phi^2}{(\Lambda'_f)^2} , ~\frac{\delta (M_V^2)}{M_V^2} = \frac{\phi^2}{(\Lambda'_V)^2} .
\end{align}
In the simplest case, in which scalar DM is produced non-thermally and forms a coherently oscillating classical field, $\phi(t) \simeq \phi_0 \cos(m_\phi t)$, which survives to the present day if the scalar particles are sufficiently light 
and weakly interacting, scalar DM that interacts with SM matter via Eq.~(\ref{Scalar_couplings_quad}) produces both `slow' linear-in-time drifts and oscillating variations of $\alpha$ and the particle masses \cite{Stadnik2015DM-VFCs}. Apart from the coherently oscillating classical fields that ultralight scalar DM can form, ultralight scalar DM may also form topological defects (TDs), which arise from the stabilisation of the scalar field under a suitable self-potential \cite{Vilenkin1985REVIEW}. In this case, TDs instead produce transient-in-time variations of $\alpha$ and the particle masses as a defect temporarily passes through some region of space \cite{Derevianko2014,Stadnik2014defects}.

\:
\:
\:

\begin{figure}[h!]
\begin{center}
\includegraphics[width=9.5cm]{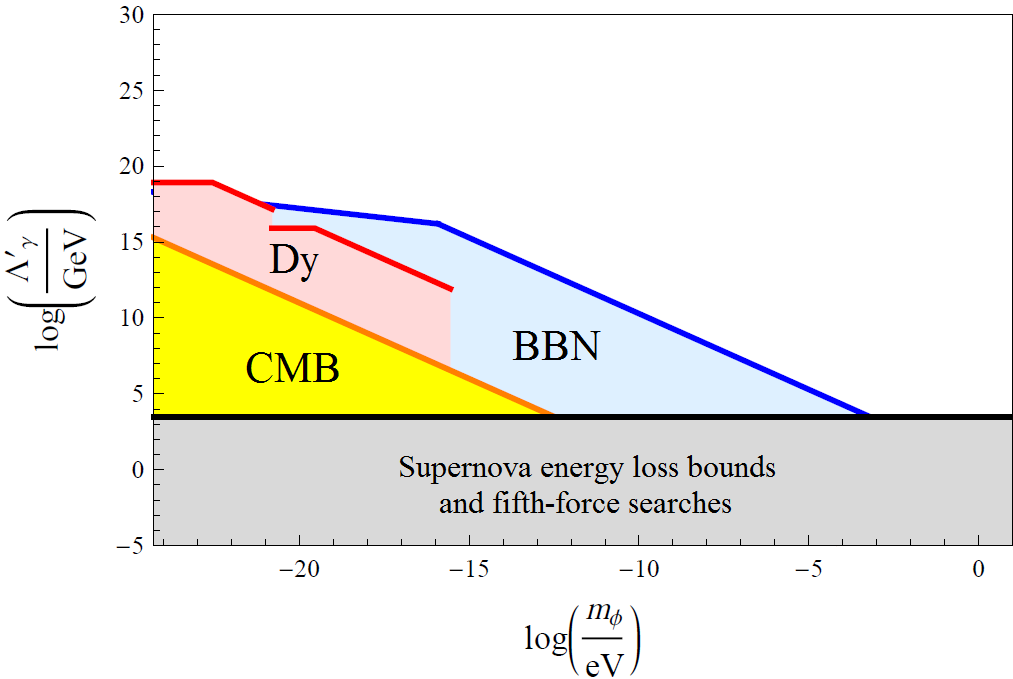}
\caption{Region of scalar dark matter parameter space ruled out for the quadratic interaction of a scalar field $\phi$ with the photon. 
Region below yellow line corresponds to constraints from CMB measurements \cite{Stadnik2015DM-VFCs}. 
Region below blue line corresponds to constraints from comparison of measurements and SM predictions of the primordial $^{4}$He abundance \cite{Stadnik2015DM-VFCs}.
Region below red line corresponds to constraints from atomic dysprosium spectroscopy measurements \cite{Stadnik2015DM-VFCs}. 
Region below black line corresponds to constraints from supernova energy loss bounds and fifth-force experimental searches \cite{Olive2008}.} 
\label{fig:Lambda_gamma_quadratic_space_+CMB}
\end{center}
\end{figure}

\textbf{BBN constraints.} ---
Astrophysical observations can be used to constrain the interactions of scalar DM with SM matter that appear in Eq.~(\ref{Scalar_couplings_quad}). Since the energy density of a non-relativistic oscillating scalar field scales as $\rho_{\textrm{scalar}} \propto (1+z)^3$, the strongest astrophysical constraints on the parameters in Eq.~(\ref{Scalar_couplings_quad}) come from the earliest observationally tested epoch of the Universe, namely Big Bang nucleosynthesis (BBN). The interactions between scalar DM and SM matter need to be sufficiently weak (see Fig.~\ref{fig:Lambda_gamma_quadratic_space_+CMB} for the case of the quadratic interaction of $\phi$ with the photon) to be consistent with the measured and predicted (within the SM) primordial $^{4}$He abundance \cite{Stadnik2015DM-VFCs}. 

\textbf{Pulsar timing searches.} ---
Astrophysical measurements can also be used to directly search for scalar DM. In particular, pulsar timing measurements can be used to search for transient-in-time variations of the neutron mass induced by TDs \cite{Stadnik2014defects}. A pulsar is a highly magnetised neutron star, which emits electromagnetic radiation and rotates with (usually a very stable) period ranging from $T \simeq 1 ~ \textrm{ms} - 10 ~ \textrm{s}$. Assuming that the angular momentum of a pulsar is conserved upon the passage of a defect through a pulsar, then its frequency of rotation would change as follows: $\delta \omega(t) / \omega \simeq -\delta m_n(t) / \delta m_n$. For sufficiently non-adiabatic passage of a defect through a pulsar, the defect may potentially trigger a `pulsar glitch' event (which have already been observed, but the exact origin of which is still disputed \cite{Haskell2015}) by triggering vortex unpinning or crustal fracture, in which the source of angular momentum required for the glitch event is provided by the pulsar itself \cite{Stadnik2015pulsar}.

\textbf{Laboratory clock searches.} ---
Transition frequencies in atomic, molecular and nuclear systems depend on the values of the fundamental constants and particle masses. Thus, comparing the ratio of two different transition frequencies, $\omega_1/\omega_2$, in such systems provides a method of searching for the oscillating-in-time and transient-in-time variations of the fundamental constants produced by scalar DM (see the reviews \cite{Flambaum2008clock_review,Flambaum2014HCI_review} for summaries of the possible systems). The first laboratory search for oscillating-in-time variation of the electromagnetic fine-structure constant has recently been performed in atomic dysprosium \cite{Budker2015scalar}, and the results have been used to place stringent constraints on the parameter $\Lambda'_\gamma$ (Fig.~\ref{fig:Lambda_gamma_quadratic_space_+CMB}) \cite{Stadnik2015DM-VFCs}.

\textbf{Laboratory laser interferometry searches.} --- 
Laser and maser interferometers may also be used to search for the oscillating-in-time and transient-in-time variations of the fundamental constants produced by scalar DM \cite{Stadnik2015laser,Stadnik2016laser}. An atomic transition frequency $\omega$ and length of a solid $L \sim N a_{\textrm{B}}$, where $a_{\textrm{B}}$ is the Bohr radius, both depend on the fundamental constants and particle masses. If the frequency of light inside an interferometer is determined by an optical atomic transition frequency and the interferometer arm length is allowed to vary freely, the atomic transition wavelength and arm length are compared directly:
\begin{align}
\label{Phase_accumulated_1}
\Phi = \frac{\omega L}{c} \propto \left(\frac{e^2}{a_{\textrm{B}} \hbar} \right) \left(\frac{N a_{\textrm{B}}}{c} \right) = N \alpha , 
\end{align}
where the optical atomic transition frequency $\omega$ is proportional to the atomic unit of frequency $e^2/a_{\textrm{B}} \hbar$.
Variation of the electromagnetic fine-structure constant thus produces the phase shift: $\delta\Phi \simeq  \Phi ~ \delta \alpha  / \alpha$.
Multiple reflections enhance observable effects due to variation of the fundamental constants by the effective mean number of passages $N_{\textrm{eff}}$.

\section*{Acknowledgments}
This work was supported by the Australian Research Council. B.~M.~R.~and V.~V.~F.~are grateful to the Mainz Institute for Theoretical Physics (MITP) for its hospitality
and support.


\begin{footnotesize}

\end{footnotesize}



\begin{thebibliography}{99}
\bibitem{Gruzinov2000fuzzyDM} W.~Hu, R.~Barkana, A.~Gruzinov, Fuzzy Cold Dark Matter:~The Wave Properties of Ultralight Particles. \emph{Phys.~Rev.~Lett.}~\textbf{85}, 1158 (2000).

\bibitem{Marsh2013DMHalo} D.~J.~E.~Marsh, J.~Silk, A model for halo formation with axion mixed dark matter. \emph{MNRAS} \textbf{437}, 2652 (2013).

\bibitem{Schive2014B} H.-Y.~Schive \emph{et al.}~Understanding the Core-Halo Relation of Quantum Wave Dark Matter, $\psi$DM, from 3D Simulations. \emph{Phys.~Rev.~Lett.}~\textbf{113}, 261302 (2014).

\bibitem{Weinberg2015ReviewDMIssues} D.~H.~Weinberg, J.~S.~Bullock, F.~Governato, R.~K.~de Naray, A.~H.~G.~Peter, Cold dark matter:~Controversies on small scales. \emph{PNAS} (2015), www.pnas.org/content/early/2015/01/27/1308716112.abstract

\bibitem{Olive2008} K.~A.~Olive, M.~Pospelov, Environmental Dependence of Masses and Coupling Constants. \emph{Phys.~Rev.~D} \textbf{77}, 043524 (2008).

\bibitem{Stadnik2015DM-VFCs} Y.~V.~Stadnik, V.~V.~Flambaum, Can Dark Matter Induce Cosmological Evolution of the Fundamental Constants of Nature? \emph{Phys.~Rev.~Lett.}~\textbf{115}, 201301 (2015).


\bibitem{Vilenkin1985REVIEW} A.~Vilenkin, Cosmic strings and domain walls. \emph{Phys.~Rep.}~\textbf{121}, 263 (1985).

\bibitem{Derevianko2014} A.~Derevianko, M.~Pospelov, Hunting for topological dark matter with atomic clocks. \emph{Nat.~Phys.}~\textbf{10}, 933 (2014).

\bibitem{Stadnik2014defects} Y.~V.~Stadnik, V.~V.~Flambaum, Searching for Topological Defect Dark Matter via Nongravitational Signatures. \emph{Phys.~Rev.~Lett.}~\textbf{113}, 151301 (2014).

\bibitem{Haskell2015} B.~Haskell, A.~Melatos, Models of Pulsar Glitches. \emph{Int.~J.~Mod.~Phys.~D} \textbf{24}, 1530008 (2015).

\bibitem{Stadnik2015pulsar} Y.~V.~Stadnik, V.~V.~Flambaum, Reply to comment on ``Searching for Topological Defect Dark Matter via Nongravitational Signatures''. arXiv:1507.01375.

\bibitem{Flambaum2008clock_review} V.~V.~Flambaum, V.~A.~Dzuba, Search for variation of the fundamental constants in atomic, molecular, and nuclear spectra. \emph{Can.~J.~Phys.}~\textbf{87}, 25 (2009).

\bibitem{Flambaum2014HCI_review} A.~Ong, J.~C.~Berengut, V.~V.~Flambaum, Highly charged ions for atomic clocks and search for variation of the fine structure constant. \emph{Springer Tracts Mod.~Phys.}~\textbf{256}, 293 (2014).

\bibitem{Budker2015scalar} K.~Van Tilburg, N.~Leefer, L.~Bougas, D.~Budker, Search for ultralight scalar dark matter with atomic spectrosopy. \emph{Phys.~Rev.~Lett.~}\textbf{115}, 011802 (2015).

\bibitem{Stadnik2015laser} Y.~V.~Stadnik, V.~V.~Flambaum, Searching for Dark Matter and Variation of Fundamental Constants with Laser and Maser Interferometry. \emph{Phys.~Rev.~Lett.}~\textbf{114}, 161301 (2015).

\bibitem{Stadnik2016laser} Y.~V.~Stadnik, V.~V.~Flambaum, Enhanced effects of variation of the fundamental constants in laser interferometers and application to dark matter detection. arXiv:1511.00447.




\end{thebibliography}
\end{document}